\documentclass[aps,prb,twocolumn]{revtex4-1}

\usepackage{amsmath}
\usepackage{bm}
\usepackage{graphicx}

\begin{document}

\title{Enhancement of thermoelectric efficiency and violation of the Wiedemann-Franz law due to
Fano effect}

\author{G. G\'omez-Silva}

\affiliation{Departamento de F\'{\i }sica, Universidad Cat\'{o}lica
del Norte, Casilla 1280, Antofagasta, Chile}

\author{O. \'Avalos-Ovando}

\affiliation{Departamento de F\'{\i }sica, Universidad Cat\'{o}lica
del Norte, Casilla 1280, Antofagasta, Chile}

\author{M. L. Ladr\'on de Guevara}
\affiliation{Departamento de F\'{\i }sica, Universidad Cat\'{o}lica
del Norte, Casilla 1280, Antofagasta, Chile}

\author{P. A. Orellana}

\affiliation{Departamento de F\'{\i }sica, Universidad Cat\'{o}lica
del Norte, Casilla 1280, Antofagasta, Chile}

\begin{abstract}
We consider the thermoelectric properties of a double-quantum-dot molecule coupled in parallel to metal electrodes with a magnetic flux threading the ring.
By means of the Sommerfeld expansion we obtain analytical expressions for the electric
and thermal conductances, thermopower and figure of merit for arbitrary values of the magnetic flux. We neglect electronic correlations. The Fano antiresonances in transmission demand that
terms usually discarded in the Sommerfeld expansion are taken into account. We also explore the behavior of the Lorenz ratio $L=\kappa/\sigma T$, where $\kappa$ and  $\sigma$ are the thermal and electrical conductances and $T$ the absolute temperature,
and we discuss the reasons why the Wiedemann-Franz law fails in presence of Fano antiresonances.
\end{abstract}

\date{\today}

\maketitle

Recently, there has been an increasing interest in the thermoelectric properties
of low dimensional systems and nanostructured materials.
Theoretical predictions\cite{hicks,khitun,balandin} as well as experiments\cite{venkatas,harman,hochbaum,boukai} show that
these structures exhibit higher efficiencies than bulk materials,\cite{thiagarajan} making them very attractive for their potential
application in energy-conversion devices.
The efficiency of a thermoelectric device is described by the dimensionless parameter $ZT$, known as figure of merit, where $T$ is the
absolute temperature and $Z$ characterizes the electrical and thermal transport properties of the device.
The figure of merit is given by $ZT=S^2\sigma T/\kappa$, where $S$, $\sigma$ and $T$ are, respectively, the thermopower (Seebeck coefficient),
electronic conductance and absolute temperature, and $\kappa=\kappa_{e}+\kappa_{ph}$ is the thermal conductance, where $\kappa_e$ is the electron
and $\kappa_{ph}$ the phonon thermal conductance.\cite{mahan} On the other hand,
the thermal and electric conductances for most macroscopic metals at very low and room temperatures are constrained by the Wiedemann-Franz law,
$\kappa/\sigma  T=L_0$, where $L_0=(\pi^3/3)(k_B/e)^2$ is the Lorenz number, with $k_B$ the Boltzmann constant and $e$
the electron charge. This relationship is a consequence of the Fermi-liquid behavior of electrons in metals, and express basically the fact that free electrons support both charge and heat transport.

While the most efficient thermoelectric bulk materials have values of $ZT$ not higher than  $1$, larger figures of merit have been measured in nanostructured
systems. Ref. \cite{venkatas} reported  $ZT=2.4$  in a thin-film superlattice device at room temperature, and $ZT=1.6$ was found in a quantum dot
superlattice.\cite{harman} More recently, two experiments showed that the thermoelectric efficiency of Si nanowires can be substantially enhanced relative to
the bulk value for Si.\cite{hochbaum,boukai}
Different mechanisms explain the improvement of thermoelectric efficiencies in systems at the nanoscale. The figure of merit can be enhanced due to the decrease of the thermal conductance produced by the scattering of phonons off the structure, \cite{hicks,khitun}
or because of the presence of an enhanced density of states at the Fermi level, which produces an increase of the thermopower.\cite{mahansofo,murphy}
Fano resonances, a signature of coherent transport of electrons, have been also predicted to improve the thermoelectric efficiency in systems as
molecules\cite{bergfield,finch,karlstrom} or multiple-level quantum dots.\cite{nakanishi} A Fano resonance arises by the quantum interference of two transport pathways, a resonant and a non-resonant one, and manifests in a characteristic asymmetric lineshape in the transmission probability.\cite{clerk}
On the other hand, since in low dimensional structures electron transport is
affected by mechanisms such as confinement, electronic correlations, and others,
the Wiedemann-Franz law is not necessarily fulfilled at the nanoscale. The violation of this law in quantum dots has been predicted at different
regimes.\cite{boese,zianni,triberis,liu3} It has been directly associated to the Coulomb interaction in a quantum dot connected to metal
leads,\cite{ahmadian,kubala} nodes in transmission\cite{bergfield} or to mesoscopic fluctuations in an open quantum dot.\cite{vavilov}

In this article we are concerned with the thermoelectric properties of a double-quantum dot molecule embedded in an Aharonov-Bohm ring.
This system is described by a transmission amplitude with two components of different spectral
linewidths, the combination between them giving rise to a
convolution of a Breit-Wigner and a Fano resonance in the transmission probability.\cite{ghost,kang}
The thermoelectricity of this system was studied numerically by Liu \emph{et al.} both in the absence\cite{liu} and in presence\cite{liu3}
of electronic correlations, finding that the figure of merit have a significant increase in the Fano lineshape regime.
Our work advances further on the findings of Refs. \cite{liu,liu3}, presenting an analytical work showing clearly that the enhancement of the
thermoelectric efficiency comes from the Fano antiresonances, which are also responsible for the failure of the Wiedemann-Franz law.
In the framework of a noninteracting model, we use the Sommerfeld expansion\cite{ashcroft} to
derive analytical expressions of the thermopower, the electric and thermal conductances, and the figure of merit.
The Fano antiresonances in transmission demand that terms usually
discarded in this expansion are taken into account.
Other analytical approaches to describe thermoelectric
effects in similar systems have been developed. Nakanishi and Kato studied the thermopower of a
multilevel quantum dot when zeros in transmission take place.\cite{nakanishi} Inasmuch as the Mott's formula, widely used for analysis of thermopower
in metals, is not valid when Fano antiresonances occur, they derived an ``extended Mott's formula" which allows an analytical calculation of this
quantity. First principles calculations of the thermoelectric efficiency of a nanojunction are developed in Ref. \cite{liu2}.

\begin{figure}[t]
  \includegraphics[width=5.4cm]{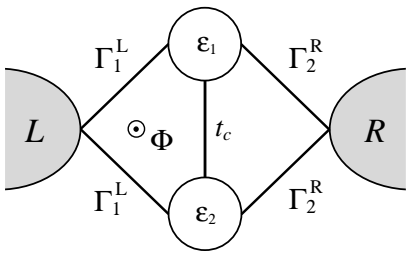}
  \caption{Scheme of the double quantum dot molecule embedded in an Aharonov-Bohm ring.}\label{fig1}
\end{figure}

The system under consideration is shown in Fig. \ref{fig1}.
Two quantum dots forming a molecule are coupled to left ($L$) and right ($R$) leads. We assume that one energy level
is relevant in the dots $1$ and $2$, with $\varepsilon_1$ and $\varepsilon_2$ their respective energies. The parameter $t_c$ is the
coupling constant between dots, and $\Gamma_{i}^{\alpha}$ the
line broadening of the energy level of the dot $i$ ($i=1,2$) due to the coupling to the lead $\alpha$ ($\alpha=L,R$),
and $\Phi$ the total magnetic flux threading the Aharonov-Bohm ring, assumed to be distributed evenly between the two sections.
The interdot and intradot electron-electron interactions are neglected.
This and similar interferometers have been carried out in two-dimensional electron gas (2DEG) systems in semiconductor heterostructures (typically AlGaAs/GaAs).\cite{kobayashi,holleitner,chen} This device has the exceptional property of being highly controllable through several of their parameters.
The energy of the discrete levels of the quantum dots can be adjusted independently via gate potentials, and so does the coupling between dots $t_c$ and the couplings between dots and leads $\Gamma_i^{L,R}$ ($i=1,2$). The phase $\phi$ is controlled through the magnetic field across the ring. The values of $\Gamma$ in these systems are of the order of a few meV.\cite{kobayashi}
The studied system also can be though as a double quantum dot molecular junction, where $\Gamma$ is of the order of $1$ eV.\cite{liu3,liu}

We model the system by means of a non-interacting Anderson Hamiltonian, as done in Ref. \cite{liu}.
We assume an effective drop voltage $\Delta V$ and a temperature difference $\Delta T$ between the left and
right leads. In the linear temperature and bias regime,
the charge current $I_e$ and the heat current $I_Q$ through the system are given
by\cite{mahanbook}
\begin{subequations}
\begin{eqnarray}
I_{e}&=&-e^2 K_0 \Delta V+ \frac{e}{T}K_1\Delta T,  \label{ie} \\
I_Q&=& e K_1 \Delta V- \frac{1}{ T}K_2\Delta T, \label{iq}
\end{eqnarray}
\label{ies}
\end{subequations}
\noindent
where $e$ is the charge of electron, $T$ the absolute temperature,  and
\begin{equation}
K_n(\mu,T)=\frac{2}{h}\int_0^\infty \left(-\frac{\partial f}{\partial E}\right) (E-\mu)^n \tau(E) \, dE,
\label{Kn}
\end{equation}
with $h$ the Planck constant, $\mu$ the Fermi energy,
$f(E-\mu )=[\exp {\beta (E-\mu )}+1]^{-1}$ the Fermi distribution, and $\beta =1/k_{B}T$, $k_{B}$ being the
Boltzmann constant, and $\tau(E)$ the transmission probability through the device.
The thermopower $S$ is defined as the voltage drop induced by a difference of temperature
when the charge current $I_e$ vanishes, and it is given by
\begin{equation}
S=-\frac{\Delta V}{\Delta T}=-\frac{1}{eT}\frac{K_{1}}{K_{0}}, \label{thermop}
\end{equation}
as follows from Eqs. (\ref{ies}). From Eq. (\ref{ie}) the electric conductance $\sigma=-I_e/\Delta V$ is obtained,
which is measured at zero temperature gradient, giving
\begin{equation}
\sigma=e^{2}K_{0}, \label{conduct}
\end{equation}
and Eqs. (\ref{ies}) lead to the electron thermal conductance $\kappa_e$, corresponding to the ratio between heat current and the temperature gradient when
the charge current is zero,
\begin{equation}
\kappa _{e}=-\frac{I_Q}{\Delta T}=\frac{1}{T}\left[ K_{2 }-\frac{%
K_{1 }^{2}}{K_{0 }}\right].  \label{thermalcond}
\end{equation}
The phonon contribution to the thermal conductance, $\kappa_{ph}$, is neglected in this model.
We assume that this has been reduced by the choice of poor thermal contacts\cite{liu2} or
by some mechanism of phonon confinement.\cite{boikov,balandin2}

The transmission $\tau(\mu)$ was obtained through the equation of motion approach
for the Green's function,\cite{eqmot} and can be written as \cite{transm}
\begin{subequations}
\label{transmiss}
\begin{equation}
\tau(\mu)=\frac{4\Gamma^2\left[t_c-(\mu-\varepsilon_0)\cos(\phi/2)\right]^2}{\Omega(\mu)},\label{trans}
\end{equation}
with
\begin{equation}
\small
\Omega(\mu)\!=\![(\mu-\varepsilon_0)^2\!-\!t_c^2\!-\Gamma^2]^2\!+\!4\Gamma^2\!\left[\mu-\!\varepsilon_0\!
-\!t_c\cos(\phi/2)\right]^2, \label{Omega}
\end{equation}
\end{subequations}
where  $\phi=2\pi \Phi/\Phi_0$ is the Aharonov-Bohm phase, $\Phi_0=h/e$ the flux quantum, and we have
assumed $\varepsilon_1=\varepsilon_2\equiv \varepsilon_0$ and $\Gamma_i^{L}=\Gamma_i^{R}\equiv \Gamma$ ($i=1,2$).
The transmission $\tau(\mu)$ is in general
a convolution of a Breit-Wigner and a Fano resonance. These resonances develop around the molecular energies
$\varepsilon_0- t_c$ and $\varepsilon_0+ t_c$, of linewidths
$2\Gamma \cos^2{(\phi/4)}$ and $2\Gamma \sin^2{(\phi/4)}$, respectively, with an antiresonance occurring at
$\mu_a=t_c \sec(\phi/2)$. The parameter $t_c$ determines how close from each other are the resonances, and depending on how large
it is as compared to $\Gamma$ the different peaks in transmission are visible or not: in the limit $t_c\ll \Gamma$ the two resonances
are not resolved, when $t_c\gg \Gamma$ they are perfectly resolved, and the case $t_c=\Gamma$ represents the crossover
between the two situations.
The phase $\phi$ determines not only the width but also the nature of each resonance,
the Fano lineshape always being the narrower one, with a width ranging from $0$ to $\Gamma$. The Breit-Wigner
resonance, in turn, has a linewidth between $\Gamma$ and $2\Gamma$.
Special features occur in the transmission spectrum when $\phi=0$ (or $2n\pi$, $n$ integer) where
only the Breit-Wigner resonance exists, the other resonance being absent due to the localization of the
associated molecular state,\cite{ghost,transm} and when $\phi=n\pi$ ($n$ odd) the transmission is a convolution
of two Breit-Wigner lineshapes of equal widths.
The transmission $\tau$ as a function of $\phi$ has a $4\pi$ period,
but its basic structure is contained in an interval of size $\pi$, where all the possible features of the
resonances are found.\cite{transm}
Given this, our analysis below only considers the interval $\phi\in [0,\pi]$.

When $\phi=0$, no Fano resonance occurs in transmission, then the quantities $\sigma$, $S$, $\kappa_e$ and $ZT$ can be
obtained resorting to the Sommerfeld expansion keeping the first non-zero term in each of the $K_0$, $K_1$ and $K_2$.
This results in  $\sigma,\kappa_e\propto \tau(\mu)$, $S\propto\tau^{(1)}(\mu)/\tau(\mu)$ (Mott's formula), and $ZT\propto 1/\tau(\mu)$.
When $0<\phi< \pi$, the transmission is a convolution of a Breit-Wigner and a Fano resonance, with a zero occurring at $\mu=\mu_a$.
Then such approximations are not valid anymore,  since both $S$ and $ZT$ diverge at that energy.
Considering terms of higher orders in $k_B T\equiv \xi$ we have
\begin{subequations} \label{sommer}
\begin{eqnarray}
K_0\!&=&\!\frac{2}{h}\left[\tau+\frac{\pi^2}{6}\tau^{(2)}\xi^2+\frac{7\pi^4}{360}\tau^{(4)}\xi^4+ O(\xi^6)\right], \label{K0} \\
K_1\!&=& \!\frac{2}{h}\left[\frac{\pi^2}{3}\tau^{(1)}\xi^2+ \frac{7\pi^4}{90}\tau^{(3)}\xi^4+O(\xi^6)\right],\label{K1}  \\
K_2\!&=& \!\frac{2}{h}\left[\frac{\pi^2}{3}\tau\xi^2+\frac{7\pi^4}{30}\tau^{(2)}\xi^4+O(\xi^6)\right], \label{K2}
\end{eqnarray} \label{kas}
\end{subequations}

\noindent
where $\tau^{(n)}\equiv\tau^{(n)}(\mu)=(d^n\tau/dE^n)(\mu)$.
It follows from Eq. (\ref{K0}) that at least the term of order $O(\xi^2)$ must be taken into account in $K_0$, in
order that the thermopower, $S\propto 1/K_0$, does not diverge at the antiresonance. On the other hand,
according to Eqs. (\ref{thermalcond}) and (\ref{kas}), $\kappa_e$ at second order vanishes at $\mu_a$, since $K_1\propto \tau^{(1)}$ and
$K_2\propto \tau$, making the figure of merit $ZT\propto 1/\kappa_e$ diverge, then terms of higher orders in $\xi$ are required.
We show below that a good agreement between the numerical and analytical curves is obtained when terms up to order $4$ are included in the expansions of the $K_n$.

In the examples below we assume $\varepsilon_0=0$, so that the two transmission resonances are located around $\mu=-t_c$ and $\mu=t_c$, and
we take $t_c=\Gamma$. We focus our attention in two values of $\phi\in (0,\pi)$ representing a narrow and a wide Fano
resonance. We consider two different values of $k_B T$, namely, $k_B T= 10^{-3}\Gamma$ and $10^{-2}\Gamma$.
For quantum dots (where $\Gamma\sim 1$ meV), these temperatures correspond to $T=10$ mK and $100$ mK, respectively, and for
molecular junctions (where $\Gamma\sim 1$ eV) to $T=10$ K and $100$ K, respectively.
\begin{figure}[ht]
  \includegraphics[width=6.2cm]{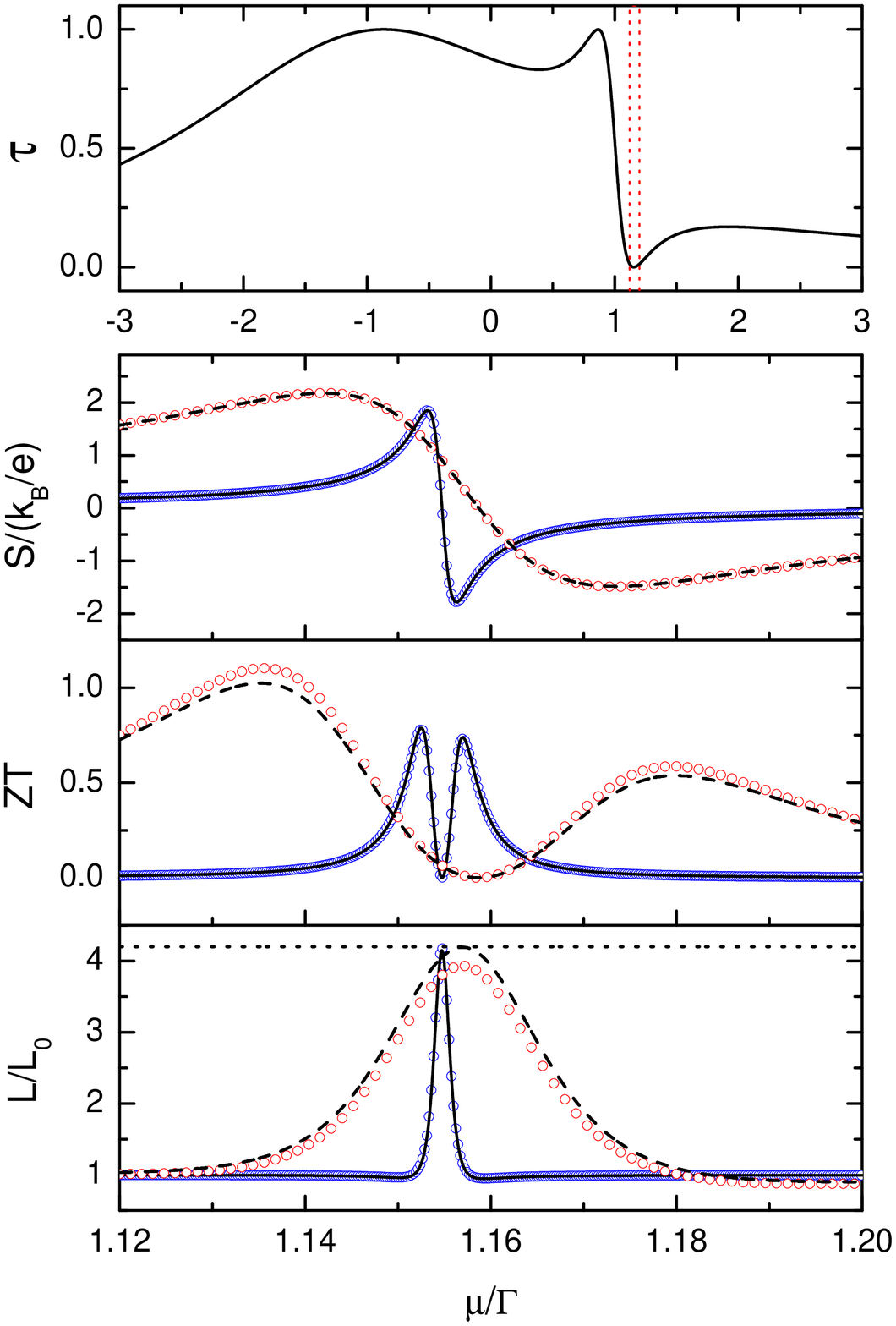}
  \caption{(Color online) Transmission probability $\tau$, thermopower $S$, figure of merit $ZT$ and ratio $L/L_0$ versus Fermi energy $\mu$ in units of $\Gamma$, for $k_BT=10^{-3}\Gamma$ (solid line, blue circles) and $k_BT=10^{-2}\Gamma$ (dash line, red circles), $\varepsilon_0=0$, $t_c=\Gamma$, and $\phi=\pi/3$. The lines stand for numerical curves, the circles for the analytical expressions via the fourth-order Sommerfeld expansion. In the upper panel we have highlighted the region of $\mu$ plotted in the lower panels. In the lowest panel we have included $L_{max}/L_0=4.19$ (dotted line). }\label{fig2}
\end{figure}

\begin{figure}[ht]
  \includegraphics[width=6.2cm]{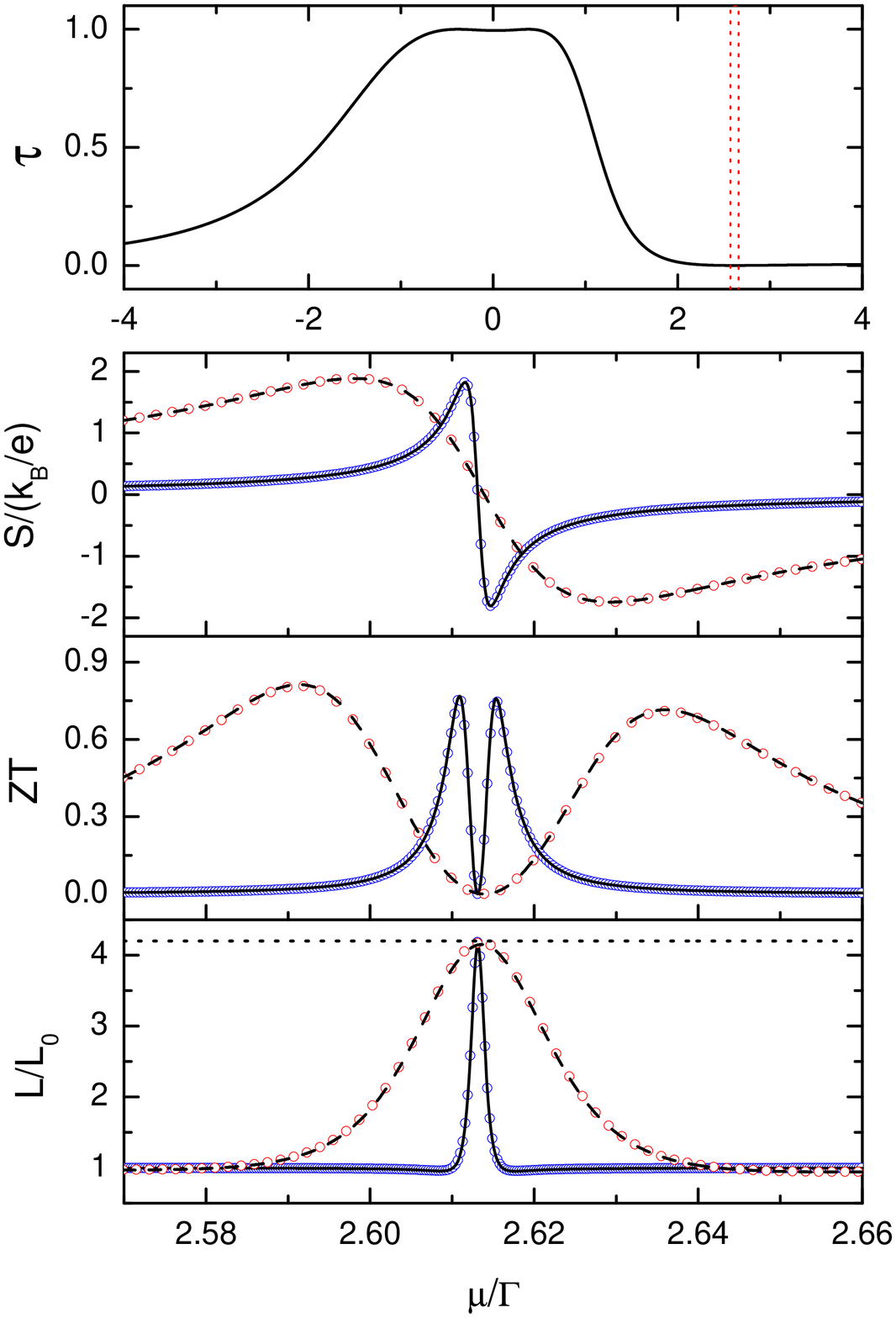}
  \caption{(Color online) Transmission probability $\tau$, thermopower $S$, figure of merit $ZT$ and ratio $L/L_0$ versus Fermi energy $\mu$ in units of $\Gamma$, for $k_B T=10^{-3}\Gamma$ (solid line, blue circles) and $k_BT=10^{-2}\Gamma$ (dash line, red circles)
  $\varepsilon_0=0$, $t_c=\Gamma$, and $\phi=3\pi/4$. The lines stand for numerical curves, the circles for the analytical expressions via the fourth-order Sommerfeld expansion. In the upper panel we have highlighted the region of $\mu$ plotted in the lower panels.
  In the lowest panel we have included $L_{max}/L_0=4.19$ (dotted line).  }\label{fig3}
\end{figure}

Figure \ref{fig2} shows the transmission $\tau$, the thermopower $S$, the figure of merit $ZT$ and the Lorenz ratio $L$ normalized by $L_0$
as a function of the Fermi energy for $\phi=\pi/3$ and two different temperatures, obtained by numerical integration of the $K_n$ (lines) and by using the Sommerfeld expansion keeping terms up to fourth order in $\xi$ (circles).
For this value of $\phi$ the transmission exhibits a Fano resonance of a linewidth much smaller than the width of the
Breit-Wigner resonance. For $k_B T=10^{-3}\Gamma$  the numerical and analytical results coincide,
while when $K_B T=10^{-2}\Gamma$ the analytical results of $ZT$ and $L/L_0$ differ slightly from the numerical ones.
We find in general that the closer to zero is $\phi$, the smaller has to be the value of $k_B T$
in order that the analytical and numerical curves match.
Figure \ref{fig3} shows analogous graphs for $\phi=3\pi/4$, where the two resonances have comparable widths.
We observe that for both values of $k_B T$, the curves obtained analytically for $S$, $ZT$ and $L$
fit exactly those obtained numerically.
For larger values of $k_B T$ the agreement between both curves is better when the Fano resonance is wider, that is, when $\phi$ approaches
to $\pi$, as evidenced by Fig. \ref{fig3}. The approximate expressions in these cases are valid even at room temperature in the case of
molecular junctions.

Now, let us briefly discuss on the role that the parameter $t_c$ has on the graphs of the thermopower and figure of merit as a
function of the Fermi energy. For $t_c=0$, that is, the two dots not forming a molecule, $S(\mu)$ is symmetric with respect to the origin,
and $ZT(\mu)$ shows peaks of equal heights. The smallest values of $ZT$ are found in this case.
If $t_c\neq 0$ the symmetries of both $S$ and $ZT$ are affected, as observed in Figs. \ref{fig2} and \ref{fig3}. If the
Fano resonance is wide enough (that is, $\phi$ is close to $\pi$),
$t_c$ does not have important effects of the shapes of both curves, just limiting to shift them horizontally, the latter being expected by the fact that $t_c$
determines the position of the Fano antiresonance. The situation changes slightly when the Fano resonance gets thinner and $\phi$ get closer
to zero. In this case the value of $t_c$ has more influence on the heights and symmetries of the two peaks present in both curves, the
highest thermoelectric efficiencies occurring when $t_c$ is close to $\Gamma$.

\begin{figure}[h]
  \includegraphics[width=4.8cm,angle=-90]{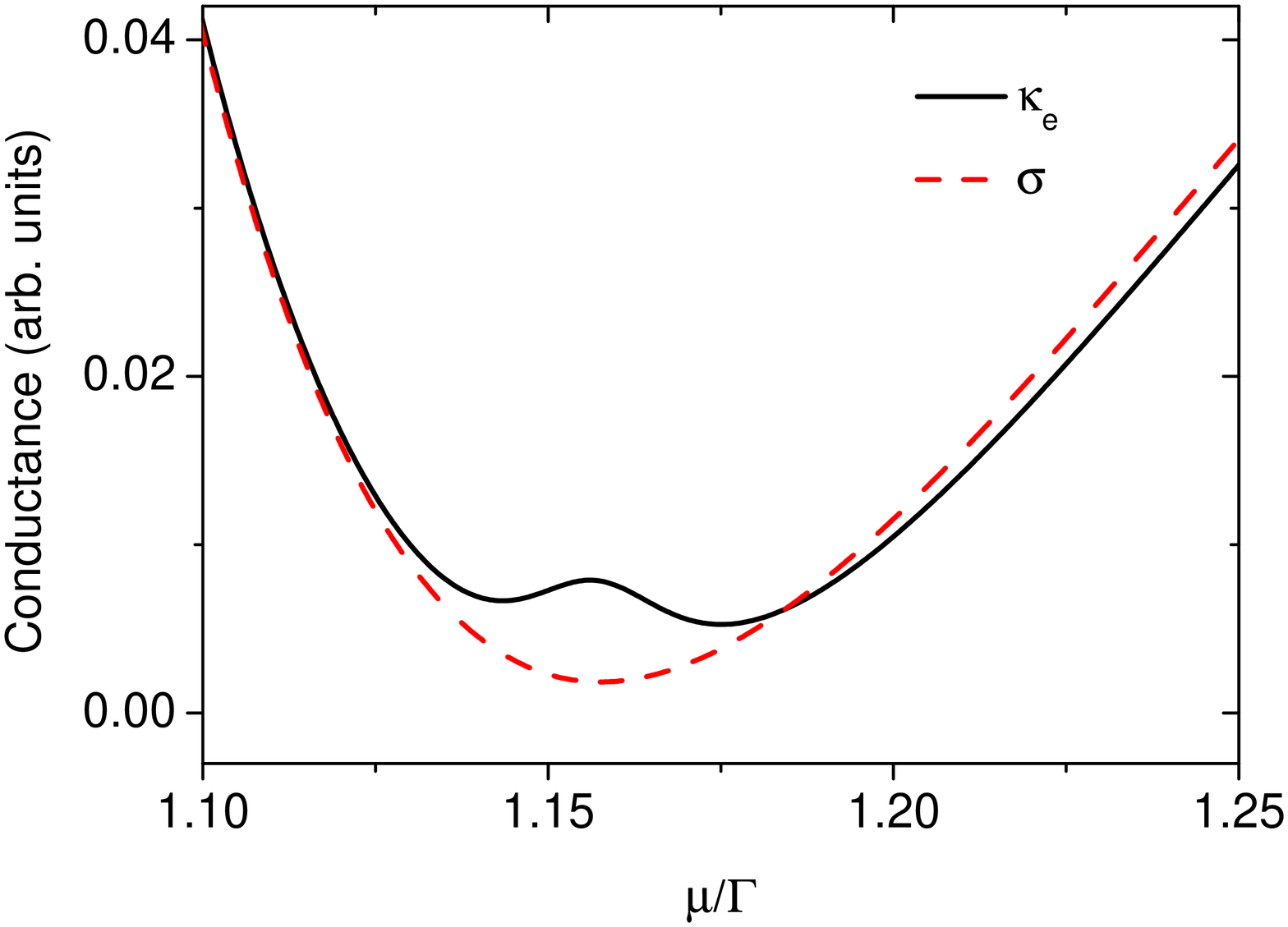}
  \caption{(Color online) Thermal and electric conductances $\kappa_e$ (solid line) and $\sigma$ (dash line) versus Fermi energy $\mu$ in units of $\Gamma$, for
  $\varepsilon_0=0$, $t_c=\Gamma$, $\phi=\pi/3$, and $k_B T=10^{-2}\Gamma$ obtained by integrating numerically Eqs. (\ref{Kn}). }\label{fig4}
\end{figure}

Last, let us pay attention to the Lorenz ratio $L$. For $\phi=0$ (and in general $\phi=2n\pi$, $n$ integer), where no Fano resonances take place,
$L=L_0$ for all values of $\mu$ at any value of $k_B T \in [0, 3\times 10^{-2} \Gamma]$. Whenever $\phi\neq 0$ ($\phi\neq 2n\pi$), as is the case of Figs.
\ref{fig2} and \ref{fig3} (lower panels), $L$ departs from $L_0$ in a small region around the antiresonance energy.
According to the expansion (\ref{K1}), $K_1$ contains only odd derivatives of the transmission
$\tau$, which is dominated by the term proportional to $\tau^{(1)}(\mu)$, which vanishes at $\mu=\mu_a$. As consequence of this, $K_1^2/K_0$ in Eq.
(\ref{thermalcond}) falls to zero close to $\mu_a$, making the thermal conductance $\kappa_e$
have a small peak in this region, while the electric conductance $\sigma$ presents a single minimum, as shown in Fig. \ref{fig4}.
The shape difference of both curves results in the violation of the Wiedemann-Franz law.
Furthermore, we observe in the examples that $L$ reaches the maximum
$L_{max}=(7 k_B^2 \pi^2) /(5e^2)=4.19 L_0$, which corresponds to the universal maximum given in Ref. \cite{bergfield}.
Although the Wiedemann-Franz law fails whenever Fano antiresonances exist, the maximum $L_{max}$ is reached for all
$\phi \neq 0$ ($2n\pi$, $n$ integer) only at very low temperatures ($k_B T\sim 10^{-3}\Gamma$ or smaller), and in general
in a constrained interval of values of $\phi$ around $\pi$, the size of this interval decreasing with temperature,
when temperatures are not very large ($k_B T$ not exceeding $2\times 10^{-2}\Gamma$).

In summary, we have used the Sommerfeld expansion to describe analytically the thermoelectric properties
of a double quantum dot molecule embedded in an Aharonov-Bohm ring, which exhibits a Fano resonance in transmission.
The existence of antiresonances demands that usually discarded terms are taken into account, in order to avoid divergences
in both the thermopower and figure of merit.
When the linewidth of the Fano resonance is close to $\Gamma$, the obtained expressions for the thermopower, figure of merit, and Lorenz ratio
are valid even at room temperature in the case of molecular junctions; for quantum dots they hold up to temperatures of the order
of tenths of Kelvin degrees.
If the Fano resonance is narrow, its linewidth being a small fraction of
$\Gamma$, these approximations hold only at very low temperatures, namely, $k_B T \sim 10^{-3} \Gamma$ or less, which
in quantum dots (molecular junctions) corresponds to temperatures of a few mK (K).
Our analysis shows clearly that the Fano antiresonances are responsible in this system for
the enhancement of the thermopower magnitude and the thermoelectric efficiency, as well as for the violation of the Wiedemann-Franz
law.

\begin{acknowledgments}
The authors acknowledge financial support from FONDECYT, under
grants 1080660 and 1100560. G. G. and O. A. thank financial
support from CONICYT Master Scholarships.
\end{acknowledgments}

\end{document}